\documentclass[nofootinbib,superscriptaddress,a4paper,twocolumn, floatfix]{revtex4-2}

\usepackage{subfigure}
\usepackage{subfloat}
\usepackage{chemfig}
\usepackage[english]{babel}
\usepackage[utf8]{inputenc}
\usepackage{blindtext}
\usepackage{amsthm}
\usepackage{mathtools}
\usepackage{physics}
\usepackage{xcolor}
\usepackage{graphicx}
\usepackage[left=23mm,right=13mm,top=35mm,columnsep=15pt]{geometry} 
\usepackage[font=small]{caption}
\usepackage{subcaption}
\usepackage{algpseudocode}
\usepackage{bm}
\usepackage{adjustbox}
\usepackage{placeins}
\usepackage[T1]{fontenc}
\usepackage{lipsum}
\usepackage{float}
\usepackage{csquotes}
\usepackage[pdftex, pdftitle={Article}, pdfauthor={Author}]{hyperref}
\usepackage{xcolor}
\hypersetup{
    colorlinks,
    linkcolor={red!50!black},
    citecolor={blue!50!black},
    urlcolor={blue!80!black}
}
\usepackage{booktabs}
\usepackage{multirow}
\usepackage{listings}
\usepackage{xcolor}

\usepackage{schemata}
\usepackage{caption}
\usepackage{subcaption}
\definecolor{codegreen}{rgb}{0,0.6,0}
\definecolor{codegray}{rgb}{0.5,0.5,0.5}
\definecolor{codepurple}{rgb}{0.58,0,0.82}
\definecolor{tqblue}{HTML}{08293d}
\definecolor{backcolour}{HTML}{fefdf5}

\lstdefinestyle{mystyle}{
    backgroundcolor=\color{backcolour},   
    commentstyle=\color{codegreen},
    keywordstyle=\color{magenta},
    numberstyle=\tiny\color{codegray},
    stringstyle=\color{codepurple},
    basicstyle=\ttfamily\footnotesize\color{tqblue},
    breakatwhitespace=false,         
    breaklines=true,
    postbreak=\mbox{\textcolor{magenta}{$\hookrightarrow$}\space},                 
    captionpos=b,                    
    keepspaces=true,                 
    numbers=left,                    
    numbersep=5pt,                  
    showspaces=false,                
    showstringspaces=false,
    showtabs=false,                  
    tabsize=2
}

\begin{document}

\title{A Hybrid Qubit Encoding:\\Splitting Fock Space into Fermionic and Bosonic Subspaces}

\author{Francisco Javier Del Arco Santos}
\email[Email:]{francisco.del.arco.santos@uni-a.de}
\affiliation{{Institute for Computer Science, University of Augsburg, Germany }}

\author{Jakob~S.~Kottmann}
\email[E-mail:]{jakob.kottmann@uni-a.de}
\affiliation{{Institute for Computer Science, University of Augsburg, Germany }}
\affiliation{{Center for Advanced Analytics and Predictive Sciences, University of Augsburg, Germany }}

\date{\today}

\begin{abstract}
Efficient encoding of electronic operators into qubits is essential for quantum chemistry simulations. The majority of methods map single electron states to qubits, effectively handling electron interactions. Alternatively, pairs of electrons can be represented as quasi-particles and encoded into qubits, significantly simplifying calculations.  This work presents a hybrid encoding that allows splitting the Fock space into Fermionic and Bosonic subspaces. By leveraging the strengths of both approaches, we provide a flexible framework for optimizing quantum simulations based on molecular characteristics and hardware constraints.
\end{abstract}

\maketitle

\section{Introduction} 
The central problem of computational chemistry is solving the Schr\"odinger Equation, which can be solved analytically for some one electron systems but usually has to be tackled numerically for many-body systems. Exact solutions can be obtained within a finite orbital basis by directly constructing and diagonalizing the Hamiltonian matrix in the basis of Slater determinants (so-called full configuration interaction or FCI) which is however only feasible for relatively small-size systems.
\\The prospect of quantum computers overcoming these temporal challenges, through direct measurement of the eigenvalues of a second-quantized operator~\cite{aspuru2005simulated}, has generated significant interest in their potential applications within the field. However, nowadays quantum computers are still being limited in the number of qubits (order of tens to a few hundred) and through their noisy nature. Furthermore, the qubits connectivity is limited and their coherence time is rather short~\cite{bharti2022noisy}.\\
As a consequence of these computational limitations, many different strategies have been developed for quantum chemistry calculations. Due to the Fermionic nature of the electrons, the total wave function must possess anti-symmetry which is incorporated through two main paradigms. On the one hand, the so-called \textit{first quantization} where the wave function is built enforcing its anti-symmetry~\cite{babbush2019Quantum}. Although the qubit requirements for these methods surpass the limitations of the current NISQ~\cite{preskill2018quantum} regime, they would apply to systems where the quantity of basis functions is significantly larger than the number of electrons~\cite{babbush2019Quantum}. 
\\On the other hand, the \textit{second quantized} representation, where the anti-symmetry is accomplished in the manner the operators are constructed. So far, many encodings have been developed in order to represent the molecule in different ways to build circuits with low depth (short length) and local (reduced number of qubits affected per action). Some of the more common ones are Jordan-Wigner (JW)~\cite{jordan1928Ueber} and  Bravyi-Kitaev (BK)~\cite{seeley2012BravyiKitaev}. More specialized encodings, focusing on compression~\cite{bravyi2017tapering, steudtner2018fermiontoqubit}, entanglement reduction~\cite{parella2024reducing} or locality and hardware constraints.~\cite{chien2020custom, derby2021compact, chien2022optimizingfermionicencodingshamiltonian, miller2023bonsai} \\
Jordan-Wigner is the most natural encoding: each spin-orbital occupation is mapped directly to one qubit, which will be in $\ket{1}$ if occupied and $\ket{0}$ otherwise. However, due to the anti-symmetry of the wave function, this encoding typically results in non-local operators and circuits since creating/annihilating the $n-$th qubit requires acting on the $n-1$ previous qubits. To solve these problems and try to perform more interesting calculations, some methods such as the Hard-Core Boson approximation have been employed~\cite{elfving2021Simulating}. Closely resembling the JW encoding, but restricting the orbital seniority (number of unpaired electrons) to zero, coupling the electrons in Bosonic quasi-particles. This results in an increase in the locality of the operators, as the quasi-particles are no longer anti-symmetric leading to improved circuit sizes and a simple grouping scheme for effective estimation of expectation values -- This however comes at the expense of generality. 
\\ Taking this into account, we have developed a flexible hybrid Fermionic-Bosonic encoding, which allows us to define each orbital encoding independently. Coupling electrons and Bosonic quasi-particles has been already applied in other fields of science,~\cite{tolle2023Exact,tolle2024G0W0}. In this context, it doesn't arise from a physical condition, it is a strong constraint that permits us to adapt the encoding to the particular case of study while saving computational resources, but without loosing the capability of naturally applying circuit designs strategies as~\cite{Izmaylov2020order, anselmetti2021local, burton2022exact, burton2024accurate, Ghasempouri2023modular}. In addition, as the electrons are now paired on the Bosonic subspace, the number of qubits is reduced from $2*N_{Mo}$ in a full Fermionic representation, to $2*N_{Fer} + N_{Bos}$.
\clearpage
\section{Basic Concepts}
\subsection{Hamiltonians}

The electronic Hamiltonian in second quantization is
\begin{equation}
\label{eq:eeham}
H=\sum_{i,j}h_{ij}\sum_{\sigma}a_{i\sigma}^\dagger a_{j\sigma} + \frac{1}{2}\sum_{i,j,k,l}g_{ij}^{kl}\sum_{\sigma\sigma'}a_{i\sigma}^\dagger a_{j\sigma'}^\dagger a_{k\sigma'}a_{l\sigma}+C,
\end{equation}
where $i,j,k,l$ goes over all the (spatial) molecular orbitals and $\sigma,\sigma' \in \{\uparrow,\downarrow\}$ represent electron spin. The integrals, here in the $[12|21]$ (or ``Google'') notation correspond to:
\begin{equation}
\label{eq:1eeint}
h_{ij}=\int\varphi_i^*(x)(-\frac{1}{2}\nabla^2-\sum_l\frac{Z_l}{r_l})\varphi_j(x)dx,
\end{equation}
\begin{equation}
\label{eq:2eeint}
g_{ij}^{kl}=\iint\frac{\varphi^*_i(x_1)\varphi^*_j(x_2)\varphi_k(x_2)\varphi_l(x_1)}{r_{12}}dx_1dx_2. 
\end{equation}
The term C corresponds to nuclear repulsion and contributions of potentially frozen occupied orbitals and will, in the following, be omitted. In our hybrid encoding, the Hamiltonian is split, restricting the occupation of certain orbitals to zero or two, coupling the electrons on Bosonic quasi-particles. Therefore, the indices on equation~\ref{eq:eeham} will be restricted to only these not-restricted (Fermionic) orbitals.
\subsection{Fermions}
To leverage a quantum processor, finding an isomorphism between the basis states of single particles, electrons in our case, and the qubits is necessary. One of the most natural and widely employed~\cite{tranter2018comparison} encodings is the 
Jordan-Wigner (JW) transformation, which we will choose in this work as well -- however, any encoding may be employed for the Fermionic part of the Hybrid encoding. The JW encoding maps fermionic annihilation (creation) operators to lowering (raising) operators on qubits
\begin{equation}  \label{eq:crea} a\propto\sigma^+_j=\frac{1}{2}(X(j)+ iY(j)).\end{equation}
These qubit operators already fulfill the annihilation (creation) properties but neglect the Fermionic anti-symmetry
\begin{subequations}
\label{eq:creanprop2}
\begin{equation}
\label{eq:anticonmutator}
    [a_i,a_j]_+\equiv a_ia_j+a_ja_i =0,
\end{equation}
\begin{equation}
\label{eq:antisymmproperties}
    [a^\dagger_i,a^\dagger_j]_+=0 \hspace{0.5cm}[a_i,a_j^\dagger]_+=\delta_{ij},
\end{equation}
\end{subequations} 
that can be incorporated through a $Z$ tensor product acting on all the previous qubits
\begin{equation}
\label{eq:antisimZ}
a_j= \left[ \prod_{n=0}^{j-1}Z(n) \right] \sigma^+(j).
\end{equation}
This incorporation of the $Z$ string is causing the non-locality of the JW encoding, where $a_N$ is forced to operate on all $N$ qubits. A unitary-coupled cluster quantum gate that implements an excitation of a single electron between spin-orbital 0 and $N$ for example has to be compiled to a relatively costly quantum circuit (see Sec.~\ref{sec:circuits}), and a similar argument holds for the time evolution generated by the encoded Hamiltonian.  
\subsection{Hard-core Bosons}
One approximation that resolves the non-locality of Fermionic encodings, while additionally reducing the number of required qubits, is spin-pairing the electrons into Bosonic quasi-particles that can occupy the spatial basis orbitals -- effectively restricting the orbital occupation to zero or two. These quasi-particles are referred to as hard-core Bosons and the corresponding Bosonic annihilation operator is
\begin{equation}
    \label{eq:JW->HCB}
    b_i \sim a_{i\uparrow}a_{i\downarrow},
\end{equation}
with the creation operator being defined accordingly. 
The hard-core Bosonic Hamiltonian can be derived from eq.~\eqref{eq:eeham}, considering only the terms which respect the hard-core Bosonic description (see for example Ref.~\cite{henderson2015Pair})
\begin{equation}
    \label{eq:h_hcb}
    H = \sum_{i,j}(g_{ii}^{jj}+2h_{ij}\delta_{ij})b_ib^\dagger_j+\sum_{i\not=j}(2g_{ji}^{ij}-g_{ji}^{ji})b^\dagger_ib_ib^\dagger_jb_j,
\end{equation}

where $b_i$~($b^\dagger_i$) is the annihilation (creation) of the hardcore Bosons acting on the $i$-th qubit. 
Since the hard-core Bosons (HCB) can occupy only spatial orbitals, we can encode them directly into qubits where qubit $k$ indicates the occupation number of spatial orbital $k$ with a hard-core Boson. The HCB  annihilation (creation) operators can be represented directly by the qubit raising and lowering operators
\begin{align}
    b_j = \sigma^+_j,\;\; b_j^\dagger = \sigma^-_j
\end{align}
\section{Hybrid-Encoding}
\label{sec:Hy-enc}
\begin{figure*}
\includegraphics[width=1.5\columnwidth]{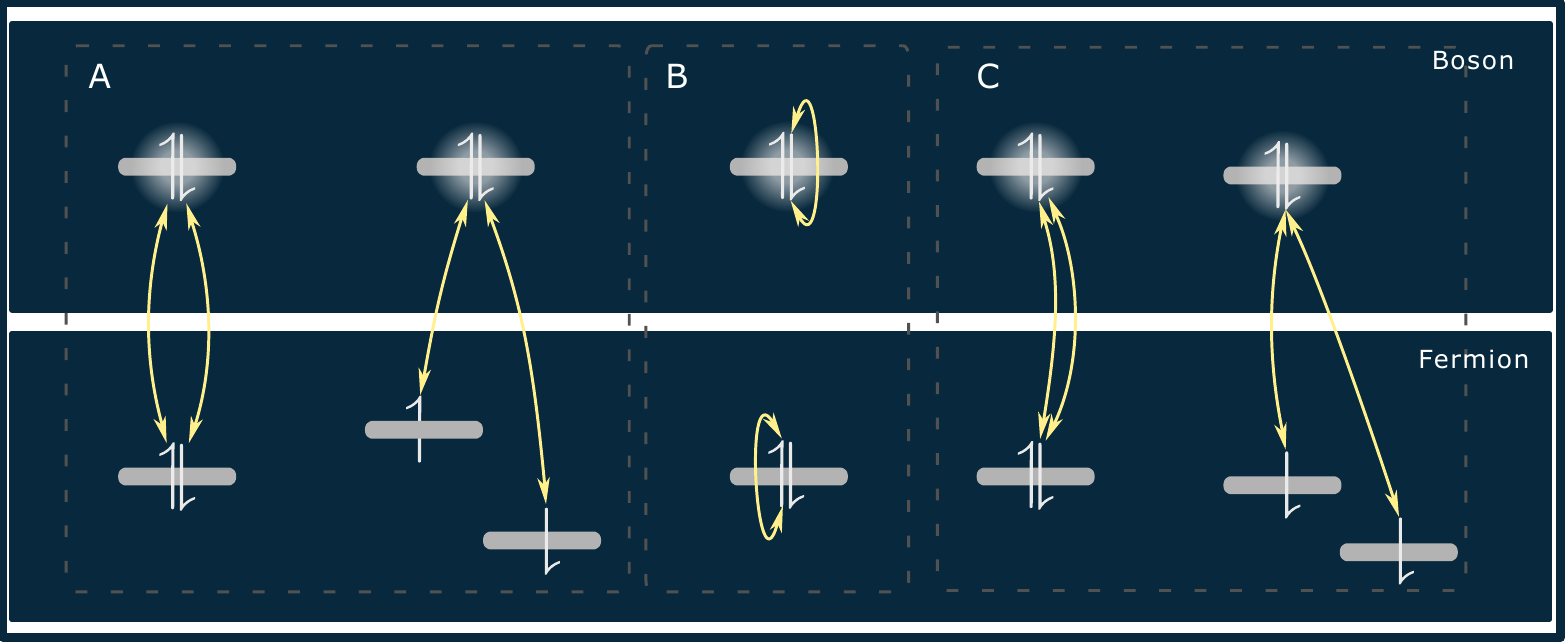}
\caption{Two-electrons interactions between the Fermionic and Bosonic subspaces.}
\label{fig:hcb_jw_interactions}
\end{figure*}
The HCB encoding typically leads to significantly shorter quantum circuits and simplified measurement schemes. It however forces all electrons in the corresponding wavefunction to form quasi-particles. Although paired-electron approximations can perform remarkably well for some systems, the obtained accuracy is often not sufficient. To balance computational needs and wavefunction accuracy we develop a hybrid encoding that splits the set of spatial orbitals into two non-overlapping sets $\mathcal{F}\cap \mathcal{B} = \emptyset$. The orbitals in the Fermionic set $\mathcal{F}$ will be fully resolved and can be occupied by spin-up and spin-down electrons while the orbitals in the Bosonic set $\mathcal{B}$ can only be occupied by a spin-paired electron pair. For this purpose, the electronic Hamiltonian will be split in three terms, one Hamiltonians over each subspace, and one interaction term
\begin{equation}
    \label{eq:H_total}
    H_{ee} = H_\text{F}+H_\text{B}+H_\text{I},
\end{equation}
with $H_\text{F}$ and $H_\text{B}$ being the usual Fermionic and HCB Hamiltonians defined on the corresponding orbital sets $\mathcal{F}$ and $\mathcal{B}$ and $H_\text{I}$ denoting an interaction operator between the two orbital sets. Consider the interaction between a spin-up and a spin-down electron in orbitals $k,l$, belonging to $\mathcal{F}$, with a quasi-particle on orbital $i$, which belongs to $\mathcal{B}$. If we want to annihilate the two resolved electrons and create a quasi-particle in orbital $i$, the second quantized operator is 
\begin{subequations}
    \begin{equation}
    \label{eq:fer_to_hi_direct}
a^\dagger_{i\uparrow}a^\dagger_{i\downarrow}a_{k\downarrow}a_{l\uparrow} \sim b^\dagger_ia_{k\downarrow}a_{l\uparrow},
\end{equation}
if we however change the ordering of the spins we receive
\begin{equation}
\label{eq:fer_to_hi_indirect}
a^\dagger_{i\downarrow}a^\dagger_{i\uparrow}a_{l\uparrow}a_{k\downarrow}\rightarrow -a^\dagger_{i\uparrow}a^\dagger_{i\downarrow}a_{l\uparrow}a_{k\downarrow}\rightarrow -b^\dagger_ia_{l\uparrow}a_{k\downarrow}.
\end{equation}
\end{subequations}
While the reordering of the spin is clearly visible in the Fermionic operators it is hidden in the Bosonic operator. This issue can be solved by fixing an ``up-then-down`` convention in the definition of the Bosonic operators (as in Eq.~\eqref{eq:JW->HCB}) and taking into account the sign when converting strings of Fermionic operators
\begin{equation}
    \label{eq:sign}
    s_{\sigma_k,\sigma_l}= \sigma_k - \sigma_l \hspace{0.3cm}\sigma_k,\sigma_l\in\{\pm\frac{1}{2}\}.
\end{equation} 
Moreover, since both groups of creation (annihilation) operators act on different subspaces, and the Bosonic spin-ordering is fixed with this sign convention, the anty-symmetry considerations, the $Z$ tensor product from eq.~\eqref{eq:antisimZ} for this JW, can be restricted to those qubits encoding Fermionic spin-orbitals. To derive the interaction Hamiltonian, only the terms that lead to the allowed occupations on the wave function should be considered, represented in Fig.~\ref{fig:hcb_jw_interactions}. We will now discuss these individual terms that enter $H_\text{I}$.\\
\begin{figure*}
\centering
\subfigure[]{
        \centering
    \includegraphics[width=0.4\textwidth]{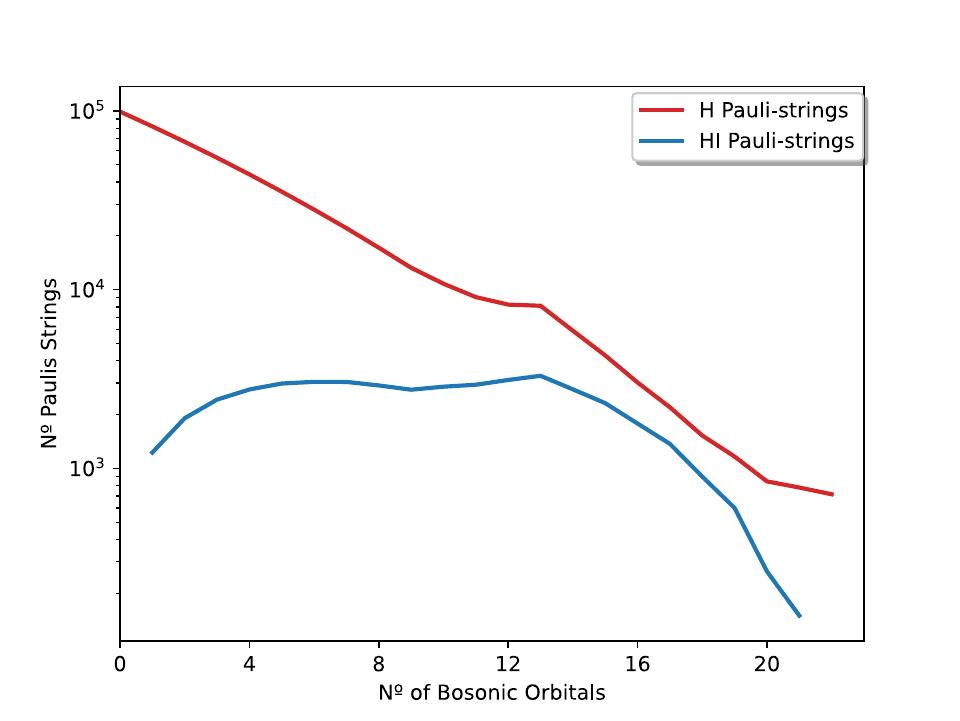}
    \label{sfig:H_paulistrings}
}
\subfigure[]{
        \centering
    \includegraphics[width=0.4\textwidth]{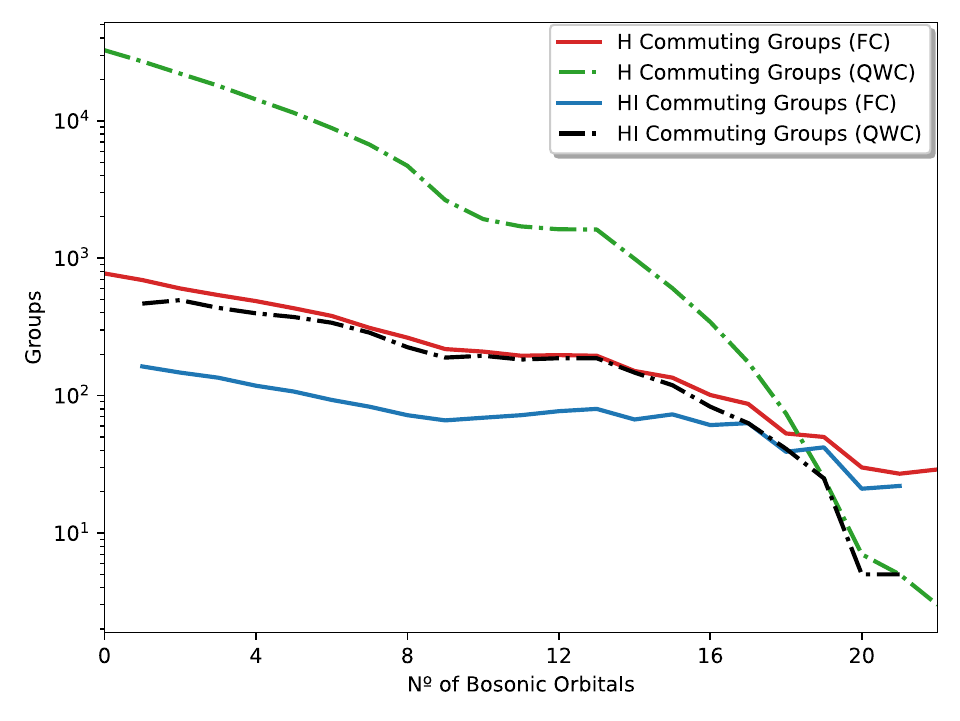}
    \label{sfig:H_conmuting} 
}
\caption{Butadiene molecule with 22 Spatial Orbitals, Fig.~\ref{sfig:H_paulistrings} presents the amount of Pauli-string for the Hybrid JW-HCB Hamiltonian as a function of the amount of Bosonic orbitals, for both the full Molecular Hamiltonian (red) and only the Interaction Hamiltonian (blue). Fig.~\ref{sfig:H_conmuting} shows the amount of commuting Pauli-string groups grouped by the Sorted Insertion method. They are grouped both by Fully Commuting (FC) condition (Red line for the Full Hamiltonian and Blue for the Interaction Terms only) and Qubit-Wise Commutativity (QWC) condition (Green line for H and Black line for $H_I$)}\label{fig:measurements}
 \end{figure*}
The first term in Fig.~\ref{fig:hcb_jw_interactions}~A describes the interaction between the electronic density where the Bosonic spatial orbital interacts with two Fermionic spin-orbitals. Taking into account equation~\ref{eq:sign}, they can be written as:
\begin{subequations}
\label{eq:HI_HHJJ_JJHH}
    \begin{equation}
    \label{eq:HI_HHJJ}
    \sum_{i\in \mathcal{B}}\sum_{k,l\in \mathcal{F}}\sum_{\sigma_k\not=\sigma_l}g^{kl}_{ii}s_{\sigma_k\sigma_l}b^\dagger_ia_{k\sigma_k}a_{l\sigma_l}
\end{equation}
\begin{equation}
    \label{eq:HI_JJHH}
    \sum_{i\in \mathcal{B}}\sum_{k,l\in \mathcal{F}}\sum_{\sigma_k\not=\sigma_l}g_{kl}^{ii}s_{\sigma_k\sigma_l}a_{k\sigma_k}^\dagger a_{l\sigma_l}^\dagger b_i
\end{equation}
\end{subequations}
Where the $i$ index corresponds to an orbital in $\mathcal{B}$, and the $k,l$ indices correspond to orbitals in $\mathcal{F}$.\\
The second family of terms, represented in Fig.~\ref{fig:hcb_jw_interactions} B, can be seen as the electron-electron coulombic repulsion.
\begin{subequations}
\label{eq:HI_JHHJ_JHHJ}
    \begin{equation}
    \label{eq:HI_JHHJ}
   \sum_{i\in \mathcal{B}}\sum_{k,l\in \mathcal{F}}\sum_{\sigma_k=\sigma_j}2g_{ki}^{il}a_{k\sigma_k}^\dagger b_{i}^\dagger   b_{i}a_{l\sigma_l} 
\end{equation}
\begin{equation}
    \label{eq:HI_HJJH}
    \sum_{i\in \mathcal{B}}\sum_{k,l\in \mathcal{F}}\sum_{\sigma_k=\sigma_l}2g_{ik}^{li}b_{i}^\dagger a_{k\sigma_k}^\dagger a_{l\sigma_l} b_{i}
\end{equation}
\end{subequations}
Where the factor of 2 accounts for the two spin combinations in the Bosonic subspace within the same $k,l$ combination.\\ 
The last terms, Fig.\ref{fig:hcb_jw_interactions} C, correspond to the exchange terms between one electron of each subspace.
\begin{subequations}
    \label{eq:HI_JHJH_HJHJ}
    \begin{equation}
    \label{eq:HI_JHJH}
   \sum_{i\in \mathcal{B}}\sum_{k,l\in \mathcal{F}}\sum_{\sigma_k=\sigma_l}g_{ki}^{li}a_{k\sigma_k}^\dagger b_{i}^\dagger   a_{l\sigma_l}b_{i}
\end{equation}
\begin{equation}
    \label{eq:HI_HJHJ}
    \sum_{i\in \mathcal{B}}\sum_{k,l\in \mathcal{F}}\sum_{\sigma_k=\sigma_l}g_{ik}^{ij}b_{i}^\dagger a_{k\sigma_k}^\dagger  b_{i}a_{l\sigma_l}
\end{equation}
\end{subequations}
The 2 factor is no longer there since only the terms with the same spin in both sub-spaces would survive. For a more compact expression, we can consider:
\begin{subequations}
    \begin{equation}
    \label{eq:pop_op_so}
    n_{i\sigma} = a^\dagger_{i\sigma}a_{i\sigma}
\end{equation}
\begin{equation}
    \label{eq:pop_op_mo}
    N_{i} = a^\dagger_{i\uparrow}a_{i\uparrow}+ a^\dagger_{i\downarrow}a_{i\downarrow}
\end{equation}
\end{subequations}
Where $n_{i\sigma}$ is the Fermionic occupation operator for a given spin-orbital, and $N_i$ the Bosonic occupation for a given spatial orbital that can also be rewritten as $N_i = b^\dagger_i b_i$. Finally, due to the operator commutation properties and these population operators, terms in eqs.~\ref{eq:HI_JHHJ_JHHJ} and \ref{eq:HI_JHJH_HJHJ} can be summarized as:
\begin{equation}
\label{eq:H_I_summed}
    \sum_{i\in \mathcal{B}}\sum_{k,l\in \mathcal{F}}\sum_{\sigma_k=\sigma_l}(2g_{ik}^{li}+2g_{ki}^{il}-g_{ki}^{lis}-g_{ik}^{il})N_{i}^\dagger a_{k\sigma_k}^\dagger a_{l\sigma_l}
\end{equation}
\subsection{Resource Reductions}
An immediate consequence of the introduced hybrid encoding is a reduction in the number of Hamiltonian terms. Without considering symmetries, the Hamiltonian is reduced from 
$\mathcal{O}(N^4)$ where $N$ is the number of orbitals to $\mathcal{O}(\abs{F}^4+\abs{B}^2+\abs{F}^2\cdot\abs{B})$.  In quantum algorithmics, the Hamiltonian terms are usually represented by Pauli-strings (tensor products of Pauli matrices). A reduction in Pauli-strings is not only advantageous for performing time evolution,~\cite{MartinezMartinez2023assessmentofvarious} but also reduce considerably the amount of measurements~\cite{yen2020measuring, verteletskyi2020measurement, choi2023measurement, bansingh2022fidelity} for the expectation values in variational approaches. When computing an expectation value $\langle H \rangle_U$ of a given circuit, it is assembled from multiple expectation values over smaller, exactly diagonalizable, operators $\langle H \rangle = \sum_k c_k \langle h_k \rangle $, with $h_k$ typically being Pauli-strings.  In the case of a pure-HCB encoding, the resulting Pauli-strings are, independent of the system's size, either of ``all-$Z$'', ``$XX$'', or ``$YY$'' type, where the individual operators in each type commute and can therefore be evaluated simultaneously -- See for example Ref.~\cite{zhao2022orbital} where this was exploited for demonstrations on ion traps. In our hybrid case, this reduction is still preserved for the Bosonic space, but for the hybrid regime (as for the Fermionic regime) we still have $X,Y,Z$ mixtures that do not commute naturally.  \\\\
In the following we will give an explicit example using the Hamiltomian of a Butadiene (C$_4$H$_6$) molecule represented with 22 Spatial Orbitals, meaning the necessary qubits will range from 22 (fully Bosonic) to 44 (fully Fermionic).\\
Figure~\ref{sfig:H_paulistrings} shows the number of Pauli-strings for both the full-Hamiltonian (red) and only the Interaction-Hamiltonian (blue) (eq.~\ref{eq:HI_HHJJ_JJHH} and~\ref{eq:H_I_summed}) for Butadiene as a function of the number of orbitals encoded with HCB while Figure~\ref{sfig:H_conmuting} shows the amount of commuting Pauli-strings estimated through the Sorted Insertion grouping method~\cite{crawford2019efficient} for both the full-Hamiltonian and the interaction term $H_I$. This is an estimation of the amount of required measurements to perform a given circuit expectation value. Both of them are presented performing the algorithm with Fully Commuting condition (FC)~\cite{yen2020measuring} and the Qubit-Wise Commuting (QWC) heuristic~\cite{verteletskyi2020measurement}.\\
We can see here, that the amount of Pauli terms and the number of commuting groups both depend on the Fermionic subspace, showing an significant reduction when increasing the Bosonic contribution. The Bosonic contributions are always many orders of magnitude under the Fermionic one, and the Interaction part only plays a significant role when approaching the full-Bosonic regime. With this we have compared the resource reduction with state-of-the art grouping methods that can be applied to the standard Hamiltonian (but also to the Fermionic and interaction Hamiltonians in the here presented hybrid mappings). 

\section{Quantum Circuits}\label{sec:circuits}
\label{subsec:ucc}
\begin{figure*}
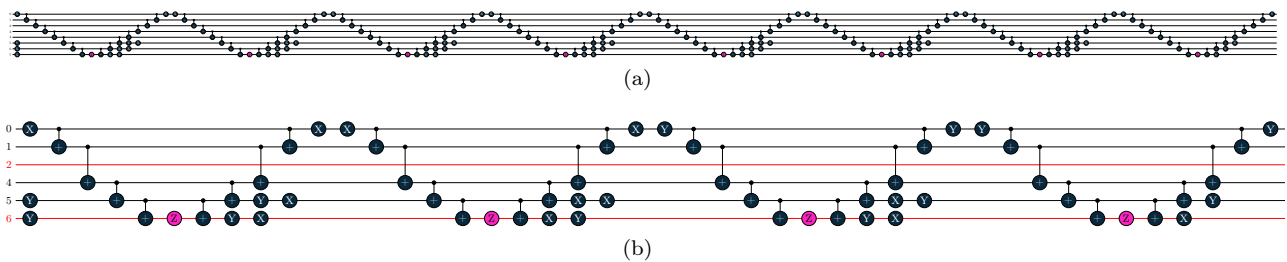

\centering
\subfigure[]{
        \centering
    \scalebox{0.145}{\input{Figures/excitation_circuits/JW_doble}}
    \label{sfig:doble_jw}
}
\subfigure[]{
        \centering
    \scalebox{0.45}{\input{Figures/excitation_circuits/Hybrid_doble}}
    \label{sfig:doble_hy}
}
\caption{Direct compilation of a two-electron excitation to single qubit Rotation and \textsc{CNOT}: \subref{sfig:doble_jw} Full JW $\phi_{0\downarrow}\phi_{2\uparrow}\rightarrow\phi_{4\downarrow}\phi_{4\uparrow}$ \subref{sfig:doble_hy} Hybrid JW-HCB of the same excitation with spatial-orbitals 0 and 2 encoded in JW, and orbitals 1 and 3 encoded in (marked with red wires) $\phi_{0\downarrow}\phi_{2\uparrow}\rightarrow\phi_{4\downarrow}\phi_{4\uparrow}$. $R_z$ gates marked in pink are carrying the gate parameter $\theta$.}
\label{fig:Compiled_Dobles}
\end{figure*}

\begin{figure*}
    \subfigure[]{
    \centering
    \scalebox{0.5}{\definecolor{tq}{rgb}{0.03137254901960784,0.1607843137254902,0.23921568627450981}
\definecolor{guo}{rgb}{0.988,0.141,0.757}
\begin{tikzpicture}[scale=1.000000,x=1pt,y=1pt]
\filldraw[color=white] (0.000000, -7.500000) rectangle (552.000000, 52.500000);
\draw[color=black] (0.000000,45.000000) -- (552.000000,45.000000);
\draw[color=black] (0.000000,45.000000) node[left] {$0$};
\draw[color=black] (0.000000,30.000000) -- (552.000000,30.000000);
\draw[color=black] (0.000000,30.000000) node[left] {$1$};
\draw[color=black] (0.000000,15.000000) -- (552.000000,15.000000);
\draw[color=black] (0.000000,15.000000) node[left] {$6$};
\draw[color=black] (0.000000,0.000000) -- (552.000000,0.000000);
\draw[color=black] (0.000000,0.000000) node[left] {$7$};
\draw (12.000000,45.000000) -- (12.000000,15.000000);
\begin{scope}
\draw[fill=tq] (12.000000, 15.000000) circle(6.000000pt);
\clip (12.000000, 15.000000) circle(6.000000pt);
\draw (12.000000, 15.000000) node {\textcolor{white}{+}};
\end{scope}
\filldraw (12.000000, 45.000000) circle(1.500000pt);
\draw (24.000000,30.000000) -- (24.000000,0.000000);
\begin{scope}
\draw[fill=tq] (24.000000, 0.000000) circle(6.000000pt);
\clip (24.000000, 0.000000) circle(6.000000pt);
\draw (24.000000, 0.000000) node {\textcolor{white}{+}};
\end{scope}
\filldraw (24.000000, 30.000000) circle(1.500000pt);
\draw (48.000000,45.000000) -- (48.000000,30.000000);
\begin{scope}
\draw[fill=tq] (48.000000, 30.000000) circle(6.000000pt);
\clip (48.000000, 30.000000) circle(6.000000pt);
\draw (48.000000, 30.000000) node {\textcolor{white}{+}};
\end{scope}
\filldraw (48.000000, 45.000000) circle(1.500000pt);
\begin{scope}
\draw[fill=tq] (48.000000, 15.000000) circle(6.000000pt);
\clip (48.000000, 15.000000) circle(6.000000pt);
\draw (48.000000, 15.000000) node {\textcolor{white}{+}};
\end{scope}
\begin{scope}
\draw[fill=tq] (48.000000, 0.000000) circle(6.000000pt);
\clip (48.000000, 0.000000) circle(6.000000pt);
\draw (48.000000, 0.000000) node {\textcolor{white}{+}};
\end{scope}
\begin{scope}
\draw[fill=tq] (72.000000, 45.000000) circle(6.000000pt);
\clip (72.000000, 45.000000) circle(6.000000pt);
\draw (72.000000, 45.000000) node {\textcolor{white}{X}};
\end{scope}
\begin{scope}
\draw[fill=guo] (96.000000, 45.000000) circle(6.000000pt);
\clip (96.000000, 45.000000) circle(6.000000pt);
\draw (96.000000, 45.000000) node {\textcolor{black}{Z}};
\end{scope}
\draw (120.000000,45.000000) -- (120.000000,15.000000);
\begin{scope}
\draw[fill=tq] (120.000000, 45.000000) circle(6.000000pt);
\clip (120.000000, 45.000000) circle(6.000000pt);
\draw (120.000000, 45.000000) node {\textcolor{white}{+}};
\end{scope}
\filldraw (120.000000, 15.000000) circle(1.500000pt);
\begin{scope}
\draw[fill=guo] (144.000000, 45.000000) circle(6.000000pt);
\clip (144.000000, 45.000000) circle(6.000000pt);
\draw (144.000000, 45.000000) node {\textcolor{black}{Z}};
\end{scope}
\draw (168.000000,45.000000) -- (168.000000,30.000000);
\begin{scope}
\draw[fill=tq] (168.000000, 45.000000) circle(6.000000pt);
\clip (168.000000, 45.000000) circle(6.000000pt);
\draw (168.000000, 45.000000) node {\textcolor{white}{+}};
\end{scope}
\filldraw (168.000000, 30.000000) circle(1.500000pt);
\begin{scope}
\draw[fill=guo] (192.000000, 45.000000) circle(6.000000pt);
\clip (192.000000, 45.000000) circle(6.000000pt);
\draw (192.000000, 45.000000) node {\textcolor{black}{Z}};
\end{scope}
\draw (216.000000,45.000000) -- (216.000000,15.000000);
\begin{scope}
\draw[fill=tq] (216.000000, 45.000000) circle(6.000000pt);
\clip (216.000000, 45.000000) circle(6.000000pt);
\draw (216.000000, 45.000000) node {\textcolor{white}{+}};
\end{scope}
\filldraw (216.000000, 15.000000) circle(1.500000pt);
\begin{scope}
\draw[fill=guo] (240.000000, 45.000000) circle(6.000000pt);
\clip (240.000000, 45.000000) circle(6.000000pt);
\draw (240.000000, 45.000000) node {\textcolor{black}{Z}};
\end{scope}
\draw (264.000000,45.000000) -- (264.000000,0.000000);
\begin{scope}
\draw[fill=tq] (264.000000, 45.000000) circle(6.000000pt);
\clip (264.000000, 45.000000) circle(6.000000pt);
\draw (264.000000, 45.000000) node {\textcolor{white}{+}};
\end{scope}
\filldraw (264.000000, 0.000000) circle(1.500000pt);
\begin{scope}
\draw[fill=guo] (288.000000, 45.000000) circle(6.000000pt);
\clip (288.000000, 45.000000) circle(6.000000pt);
\draw (288.000000, 45.000000) node {\textcolor{black}{Z}};
\end{scope}
\draw (312.000000,45.000000) -- (312.000000,15.000000);
\begin{scope}
\draw[fill=tq] (312.000000, 45.000000) circle(6.000000pt);
\clip (312.000000, 45.000000) circle(6.000000pt);
\draw (312.000000, 45.000000) node {\textcolor{white}{+}};
\end{scope}
\filldraw (312.000000, 15.000000) circle(1.500000pt);
\begin{scope}
\draw[fill=guo] (336.000000, 45.000000) circle(6.000000pt);
\clip (336.000000, 45.000000) circle(6.000000pt);
\draw (336.000000, 45.000000) node {\textcolor{black}{Z}};
\end{scope}
\draw (360.000000,45.000000) -- (360.000000,30.000000);
\begin{scope}
\draw[fill=tq] (360.000000, 45.000000) circle(6.000000pt);
\clip (360.000000, 45.000000) circle(6.000000pt);
\draw (360.000000, 45.000000) node {\textcolor{white}{+}};
\end{scope}
\filldraw (360.000000, 30.000000) circle(1.500000pt);
\begin{scope}
\draw[fill=guo] (384.000000, 45.000000) circle(6.000000pt);
\clip (384.000000, 45.000000) circle(6.000000pt);
\draw (384.000000, 45.000000) node {\textcolor{black}{Z}};
\end{scope}
\draw (408.000000,45.000000) -- (408.000000,15.000000);
\begin{scope}
\draw[fill=tq] (408.000000, 45.000000) circle(6.000000pt);
\clip (408.000000, 45.000000) circle(6.000000pt);
\draw (408.000000, 45.000000) node {\textcolor{white}{+}};
\end{scope}
\filldraw (408.000000, 15.000000) circle(1.500000pt);
\begin{scope}
\draw[fill=guo] (432.000000, 45.000000) circle(6.000000pt);
\clip (432.000000, 45.000000) circle(6.000000pt);
\draw (432.000000, 45.000000) node {\textcolor{black}{Z}};
\end{scope}
\begin{scope}
\draw[fill=tq] (432.000000, 15.000000) circle(6.000000pt);
\clip (432.000000, 15.000000) circle(6.000000pt);
\draw (432.000000, 15.000000) node {\textcolor{white}{+}};
\end{scope}
\draw (456.000000,45.000000) -- (456.000000,0.000000);
\begin{scope}
\draw[fill=tq] (456.000000, 45.000000) circle(6.000000pt);
\clip (456.000000, 45.000000) circle(6.000000pt);
\draw (456.000000, 45.000000) node {\textcolor{white}{+}};
\end{scope}
\filldraw (456.000000, 0.000000) circle(1.500000pt);
\begin{scope}
\draw[fill=tq] (480.000000, 45.000000) circle(6.000000pt);
\clip (480.000000, 45.000000) circle(6.000000pt);
\draw (480.000000, 45.000000) node {\textcolor{white}{X}};
\end{scope}
\begin{scope}
\draw[fill=tq] (480.000000, 0.000000) circle(6.000000pt);
\clip (480.000000, 0.000000) circle(6.000000pt);
\draw (480.000000, 0.000000) node {\textcolor{white}{+}};
\end{scope}
\draw (504.000000,45.000000) -- (504.000000,30.000000);
\begin{scope}
\draw[fill=tq] (504.000000, 30.000000) circle(6.000000pt);
\clip (504.000000, 30.000000) circle(6.000000pt);
\draw (504.000000, 30.000000) node {\textcolor{white}{+}};
\end{scope}
\filldraw (504.000000, 45.000000) circle(1.500000pt);
\draw (528.000000,30.000000) -- (528.000000,0.000000);
\begin{scope}
\draw[fill=tq] (528.000000, 0.000000) circle(6.000000pt);
\clip (528.000000, 0.000000) circle(6.000000pt);
\draw (528.000000, 0.000000) node {\textcolor{white}{+}};
\end{scope}
\filldraw (528.000000, 30.000000) circle(1.500000pt);
\draw (540.000000,45.000000) -- (540.000000,15.000000);
\begin{scope}
\draw[fill=tq] (540.000000, 15.000000) circle(6.000000pt);
\clip (540.000000, 15.000000) circle(6.000000pt);
\draw (540.000000, 15.000000) node {\textcolor{white}{+}};
\end{scope}
\filldraw (540.000000, 45.000000) circle(1.500000pt);
\end{tikzpicture}}
    \label{sfig:paired_jw}
}
    \subfigure[]{
    \centering
    \scalebox{0.6}{\definecolor{tq}{rgb}{0.03137254901960784,0.1607843137254902,0.23921568627450981}
\definecolor{guo}{rgb}{0.988,0.141,0.757}
\begin{tikzpicture}[scale=1.000000,x=1pt,y=1pt]
\filldraw[color=white] (0.000000, -7.500000) rectangle (192.000000, 22.500000);
\draw[color=red] (0.000000,15.000000) -- (192.000000,15.000000);
\draw[color=red] (0.000000,15.000000) node[left] {$0$};
\draw[color=red] (0.000000,0.000000) -- (192.000000,0.000000);
\draw[color=red] (0.000000,0.000000) node[left] {$3$};
\draw (12.000000,15.000000) -- (12.000000,0.000000);
\begin{scope}
\draw[fill=tq] (12.000000, 15.000000) circle(6.000000pt);
\clip (12.000000, 15.000000) circle(6.000000pt);
\draw (12.000000, 15.000000) node {\textcolor{white}{+}};
\end{scope}
\filldraw (12.000000, 0.000000) circle(1.500000pt);
\begin{scope}
\draw[fill=tq] (36.000000, 0.000000) circle(6.000000pt);
\clip (36.000000, 0.000000) circle(6.000000pt);
\draw (36.000000, 0.000000) node {\textcolor{white}{X}};
\end{scope}
\begin{scope}
\draw[fill=guo] (60.000000, 0.000000) circle(6.000000pt);
\clip (60.000000, 0.000000) circle(6.000000pt);
\draw (60.000000, 0.000000) node {\textcolor{black}{Z}};
\end{scope}
\draw (84.000000,15.000000) -- (84.000000,0.000000);
\begin{scope}
\draw[fill=tq] (84.000000, 0.000000) circle(6.000000pt);
\clip (84.000000, 0.000000) circle(6.000000pt);
\draw (84.000000, 0.000000) node {\textcolor{white}{+}};
\end{scope}
\filldraw (84.000000, 15.000000) circle(1.500000pt);
\begin{scope}
\draw[fill=guo] (108.000000, 0.000000) circle(6.000000pt);
\clip (108.000000, 0.000000) circle(6.000000pt);
\draw (108.000000, 0.000000) node {\textcolor{black}{Z}};
\end{scope}
\draw (132.000000,15.000000) -- (132.000000,0.000000);
\begin{scope}
\draw[fill=tq] (132.000000, 0.000000) circle(6.000000pt);
\clip (132.000000, 0.000000) circle(6.000000pt);
\draw (132.000000, 0.000000) node {\textcolor{white}{+}};
\end{scope}
\filldraw (132.000000, 15.000000) circle(1.500000pt);
\begin{scope}
\draw[fill=tq] (156.000000, 0.000000) circle(6.000000pt);
\clip (156.000000, 0.000000) circle(6.000000pt);
\draw (156.000000, 0.000000) node {\textcolor{white}{X}};
\end{scope}
\draw (180.000000,15.000000) -- (180.000000,0.000000);
\begin{scope}
\draw[fill=tq] (180.000000, 15.000000) circle(6.000000pt);
\clip (180.000000, 15.000000) circle(6.000000pt);
\draw (180.000000, 15.000000) node {\textcolor{white}{+}};
\end{scope}
\filldraw (180.000000, 0.000000) circle(1.500000pt);
\end{tikzpicture}}
    \label{sfig:opt_hcb}
}
\caption{Optimized compilation of a paired double excitation $\phi_{0\downarrow}\phi_{0\uparrow}\rightarrow\phi_{3\downarrow}\phi_{3\uparrow}$. \subref{sfig:paired_jw} JW encoding. \subref{sfig:opt_hcb} HCB encoding. $R_z$ gates marked in pink are carrying the gate parameter $\theta$.}\label{fig:Compiled_paired_Dobles_Opt}
\end{figure*}
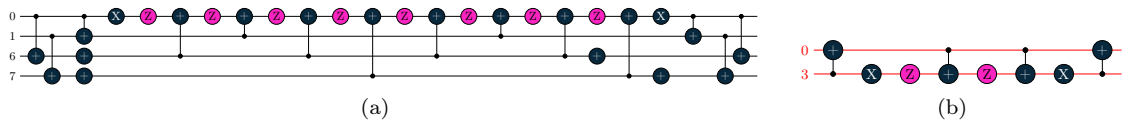

Coming from classical quantum chemistry, Unitary Coupled Cluster (UCC)~\cite{anand2022quantum} is the main physically driven circuit design framework that was already used in the first variational quantum algorithms.~\cite{mcclean2016theory, romero2018strategies} The core idea is to generate a whole group of excitations via: 
\begin{equation}
    \label{eq:U_UCC}
    U(\theta)=e^{i\frac{\theta}{2}(\hat{T}-\hat{T}^\dagger)}=e^{\sum_j i\frac{\theta_j}{2}(\hat{T}_j-\hat{T}_j^\dagger)}
\end{equation}
Where $\hat{T}$ ($\hat{T}^\dagger$) are the excitation (de-excitation) generators, defined as:
\begin{subequations}\label{eq:exct}
    \begin{equation}
    \label{eq:T_operator}
    \hat{T}=\hat{T}_1+\hat{T}_2+\hat{T}_3 + ...
\end{equation}
\begin{equation}
    \label{eq:T_n_operator}
    \hat{T}_n = \frac{1}{(n!)^2}\sum_{\substack{i,j,..\\a,b,...}}\theta_{i,j,...}^{a,b,...}a_a^\dagger a_b^\dagger... a_i a_j...
\end{equation}
\end{subequations}
To each excitation $T$, its respective complex conjugate is subtracted, the "de-excitation", making this exponential a unitary operator. $\theta_j$  is the variational parameter associated with the j-th excitation which will be optimized on the Variational Quantum Eigensolver algorithm, by minimizing the Hamiltonian expectation value,
\begin{equation}
    \label{eq:VQE_E}
    E_{min} = \min_{\theta}\langle\hat{H}\rangle_{\hat{U}(\theta)}
\end{equation}
\begin{equation}
    \label{eq:VQE_EXP}
    \langle\hat{H}\rangle_{\hat{U}(\theta)}\equiv\bra{\Psi_0}\hat{U}^\dagger(\theta)\hat{H}\hat{U}(\theta)\ket{\Psi_0}
\end{equation}
where $\ket{\Psi_0}$ is the reference state (typically the Hartree-Fock state). Traditionally, all combinations of the excitation indexes are taken into account, however, it would rapidly increase the number of terms as a function of the basis set. For this reason, Paired CC Doubles (pCCD) has been developed, where only the double excitations that move electrons-pairs are allowed. It was tested originally on classical CC, which already proved to include strong correlation~\cite{henderson2014seniority,henderson2015Pair}, describing correctly single bond breaking, but fails in multiple bond breaking. It has been translated to UCC (UpCCSD) including all the regular single excitations, and was shown to produce accurate electronic energies on test systems.~\cite{lee2018generalized}. \\ 
In practice, these methods are implemented by splitting eq.~\ref{eq:U_UCC} (Trotterization),
\begin{equation}
    \label{eq:U_UCC_trotter}
    U(\theta)=e^{\sum_j i\frac{\theta_j}{2}(\hat{T}_j-\hat{T}_j^\dagger)}\approx\prod_{i,j}e^{i\frac{\theta_{i,j}}{2}(\hat{T}_{i,j}-\hat{T}_{i,j}^\dagger)},
\end{equation}
leading to a circuit built by layers of n-electron excitations. In this job, the implementation of these building circuits has been affected in two ways. To adapt these circuit designs to the here proposed hybrid-encoding, we modify the excitation gates compilation according to the transitions described in Sec.~\ref{sec:Hy-enc}. Additionally, all excitations that violate the occupation restriction have been removed, including one-electron excitations acting on the Bosonic subspace. Two-electron excitations are only retained if they adhere to the schemas in Figure~\ref{fig:hcb_jw_interactions}. \\\\
To illustrate this compilation reduction, Figures~\ref{fig:Compiled_Dobles}, ~\ref{fig:Compiled_Dobles_Opt}, and ~\ref{fig:Compiled_paired_Dobles_Opt} present the compilation to \textsc{CNOT}s and single qubit rotations in standard JW and with the proposed hybrid encoding. Compilation of a two-electron semi-paired excitation,  which promotes two electrons from different orbitals into the same one, is shown in Figs.~\ref{fig:Compiled_Dobles_Opt} where we used an optimized optimized decomposition following the procedure described on \cite{yordanov2020efficient} and ~\ref{fig:Compiled_Dobles} showing the direct compilation of the associated Pauli strings, serving as a stand-in for encodings without optimized compilation. In Fig.~\ref{fig:Compiled_paired_Dobles_Opt} we show the same approach for a fully-paired double excitation which can therefore be encoded in full HCB if the orbitals are part of $\mathcal{B}$.\\
These results present the qubit reduction and the decrease in the depth of these circuits by more than a factor of two, independent of compilation scheme.  These semi-paired excitations have been presented since they clearly show the main compiling reduction advantages of this hybrid encoding. Still, these advantages will be present in any further excitation in this schema.

 \begin{figure*}

\subfigure[]{
        \centering
    \scalebox{0.4}{\input{Figures/excitation_circuits/compiled_Fermionic}}
    \label{sfig:opt_jw}
}
\subfigure[]{
        \centering
    \scalebox{0.45}{\definecolor{tq}{rgb}{0.03137254901960784,0.1607843137254902,0.23921568627450981}
\definecolor{guo}{rgb}{0.988,0.141,0.757}
\begin{tikzpicture}[scale=1.000000,x=1pt,y=1pt]
\filldraw[color=white] (0.000000, -7.500000) rectangle (372.000000, 67.500000);
\draw[color=black] (0.000000,60.000000) -- (372.000000,60.000000);
\draw[color=black] (0.000000,60.000000) node[left] {$0$};
\draw[color=black] (0.000000,45.000000) -- (372.000000,45.000000);
\draw[color=black] (0.000000,45.000000) node[left] {$1$};
\draw[color=black] (0.000000,30.000000) -- (372.000000,30.000000);
\draw[color=black] (0.000000,30.000000) node[left] {$4$};
\draw[color=black] (0.000000,15.000000) -- (372.000000,15.000000);
\draw[color=black] (0.000000,15.000000) node[left] {$5$};
\draw[color=red] (0.000000,0.000000) -- (372.000000,0.000000);
\draw[color=red] (0.000000,0.000000) node[left] {$6$};
\draw (12.000000,60.000000) -- (12.000000,15.000000);
\begin{scope}
\draw[fill=tq] (12.000000, 60.000000) circle(6.000000pt);
\clip (12.000000, 60.000000) circle(6.000000pt);
\draw (12.000000, 60.000000) node {\textcolor{white}{+}};
\end{scope}
\filldraw (12.000000, 15.000000) circle(1.500000pt);
\draw (24.000000,45.000000) -- (24.000000,30.000000);
\begin{scope}
\draw[fill=tq] (24.000000, 45.000000) circle(6.000000pt);
\clip (24.000000, 45.000000) circle(6.000000pt);
\draw (24.000000, 45.000000) node {\textcolor{white}{+}};
\end{scope}
\filldraw (24.000000, 30.000000) circle(1.500000pt);
\draw (48.000000,15.000000) -- (48.000000,0.000000);
\begin{scope}
\draw[fill=tq] (48.000000, 0.000000) circle(6.000000pt);
\clip (48.000000, 0.000000) circle(6.000000pt);
\draw (48.000000, 0.000000) node {\textcolor{white}{+}};
\end{scope}
\filldraw (48.000000, 15.000000) circle(1.500000pt);
\begin{scope}
\draw[fill=tq] (48.000000, 60.000000) circle(6.000000pt);
\clip (48.000000, 60.000000) circle(6.000000pt);
\draw (48.000000, 60.000000) node {\textcolor{white}{+}};
\end{scope}
\draw (72.000000,45.000000) -- (72.000000,15.000000);
\begin{scope}
\draw[fill=tq] (72.000000, 15.000000) +(-45.000000:8.485281pt and 8.485281pt) -- +(45.000000:8.485281pt and 8.485281pt) -- +(135.000000:8.485281pt and 8.485281pt) -- +(225.000000:8.485281pt and 8.485281pt) -- cycle;
\clip (72.000000, 15.000000) +(-45.000000:8.485281pt and 8.485281pt) -- +(45.000000:8.485281pt and 8.485281pt) -- +(135.000000:8.485281pt and 8.485281pt) -- +(225.000000:8.485281pt and 8.485281pt) -- cycle;
\draw (72.000000, 15.000000) node {\textcolor{white}{Z}};
\end{scope}
\filldraw (72.000000, 45.000000) circle(1.500000pt);
\begin{scope}
\draw[fill=tq] (72.000000, 60.000000) +(-45.000000:8.485281pt and 8.485281pt) -- +(45.000000:8.485281pt and 8.485281pt) -- +(135.000000:8.485281pt and 8.485281pt) -- +(225.000000:8.485281pt and 8.485281pt) -- cycle;
\clip (72.000000, 60.000000) +(-45.000000:8.485281pt and 8.485281pt) -- +(45.000000:8.485281pt and 8.485281pt) -- +(135.000000:8.485281pt and 8.485281pt) -- +(225.000000:8.485281pt and 8.485281pt) -- cycle;
\draw (72.000000, 60.000000) node {\textcolor{white}{H}};
\end{scope}
\begin{scope}
\draw[fill=tq] (72.000000, -0.000000) +(-45.000000:8.485281pt and 8.485281pt) -- +(45.000000:8.485281pt and 8.485281pt) -- +(135.000000:8.485281pt and 8.485281pt) -- +(225.000000:8.485281pt and 8.485281pt) -- cycle;
\clip (72.000000, -0.000000) +(-45.000000:8.485281pt and 8.485281pt) -- +(45.000000:8.485281pt and 8.485281pt) -- +(135.000000:8.485281pt and 8.485281pt) -- +(225.000000:8.485281pt and 8.485281pt) -- cycle;
\draw (72.000000, -0.000000) node {\textcolor{white}{H}};
\end{scope}
\begin{scope}
\draw[fill=guo] (96.000000, 15.000000) circle(6.000000pt);
\clip (96.000000, 15.000000) circle(6.000000pt);
\draw (96.000000, 15.000000) node {\textcolor{black}{Y}};
\end{scope}
\draw (120.000000,60.000000) -- (120.000000,15.000000);
\begin{scope}
\draw[fill=tq] (120.000000, 60.000000) circle(6.000000pt);
\clip (120.000000, 60.000000) circle(6.000000pt);
\draw (120.000000, 60.000000) node {\textcolor{white}{+}};
\end{scope}
\filldraw (120.000000, 15.000000) circle(1.500000pt);
\begin{scope}
\draw[fill=guo] (144.000000, 15.000000) circle(6.000000pt);
\clip (144.000000, 15.000000) circle(6.000000pt);
\draw (144.000000, 15.000000) node {\textcolor{black}{Y}};
\end{scope}
\draw (168.000000,15.000000) -- (168.000000,0.000000);
\begin{scope}
\draw[fill=tq] (168.000000, 0.000000) circle(6.000000pt);
\clip (168.000000, 0.000000) circle(6.000000pt);
\draw (168.000000, 0.000000) node {\textcolor{white}{+}};
\end{scope}
\filldraw (168.000000, 15.000000) circle(1.500000pt);
\begin{scope}
\draw[fill=guo] (192.000000, 15.000000) circle(6.000000pt);
\clip (192.000000, 15.000000) circle(6.000000pt);
\draw (192.000000, 15.000000) node {\textcolor{black}{Y}};
\end{scope}
\begin{scope}
\draw[fill=tq] (192.000000, 0.000000) circle(6.000000pt);
\clip (192.000000, 0.000000) circle(6.000000pt);
\draw (192.000000, 0.000000) node {\textcolor{white}{Z}};
\end{scope}
\draw (216.000000,60.000000) -- (216.000000,15.000000);
\begin{scope}
\draw[fill=tq] (216.000000, 60.000000) circle(6.000000pt);
\clip (216.000000, 60.000000) circle(6.000000pt);
\draw (216.000000, 60.000000) node {\textcolor{white}{+}};
\end{scope}
\filldraw (216.000000, 15.000000) circle(1.500000pt);
\begin{scope}
\draw[fill=guo] (240.000000, 15.000000) circle(6.000000pt);
\clip (240.000000, 15.000000) circle(6.000000pt);
\draw (240.000000, 15.000000) node {\textcolor{black}{Y}};
\end{scope}
\begin{scope}
\draw[fill=tq] (240.000000, 60.000000) +(-45.000000:8.485281pt and 8.485281pt) -- +(45.000000:8.485281pt and 8.485281pt) -- +(135.000000:8.485281pt and 8.485281pt) -- +(225.000000:8.485281pt and 8.485281pt) -- cycle;
\clip (240.000000, 60.000000) +(-45.000000:8.485281pt and 8.485281pt) -- +(45.000000:8.485281pt and 8.485281pt) -- +(135.000000:8.485281pt and 8.485281pt) -- +(225.000000:8.485281pt and 8.485281pt) -- cycle;
\draw (240.000000, 60.000000) node {\textcolor{white}{H}};
\end{scope}
\draw (264.000000,45.000000) -- (264.000000,15.000000);
\begin{scope}
\draw[fill=tq] (264.000000, 15.000000) +(-45.000000:8.485281pt and 8.485281pt) -- +(45.000000:8.485281pt and 8.485281pt) -- +(135.000000:8.485281pt and 8.485281pt) -- +(225.000000:8.485281pt and 8.485281pt) -- cycle;
\clip (264.000000, 15.000000) +(-45.000000:8.485281pt and 8.485281pt) -- +(45.000000:8.485281pt and 8.485281pt) -- +(135.000000:8.485281pt and 8.485281pt) -- +(225.000000:8.485281pt and 8.485281pt) -- cycle;
\draw (264.000000, 15.000000) node {\textcolor{white}{Z}};
\end{scope}
\filldraw (264.000000, 45.000000) circle(1.500000pt);
\begin{scope}
\draw[fill=tq] (264.000000, 60.000000) circle(6.000000pt);
\clip (264.000000, 60.000000) circle(6.000000pt);
\draw (264.000000, 60.000000) node {\textcolor{white}{+}};
\end{scope}
\draw (288.000000,45.000000) -- (288.000000,30.000000);
\begin{scope}
\draw[fill=tq] (288.000000, 45.000000) circle(6.000000pt);
\clip (288.000000, 45.000000) circle(6.000000pt);
\draw (288.000000, 45.000000) node {\textcolor{white}{+}};
\end{scope}
\filldraw (288.000000, 30.000000) circle(1.500000pt);
\begin{scope}
\draw[fill=tq] (288.000000, 15.000000) circle(6.000000pt);
\clip (288.000000, 15.000000) circle(6.000000pt);
\draw (288.000000, 15.000000) node {\textcolor{white}{Z}};
\end{scope}
\draw (312.000000,15.000000) -- (312.000000,0.000000);
\begin{scope}
\draw[fill=tq] (312.000000, 0.000000) circle(6.000000pt);
\clip (312.000000, 0.000000) circle(6.000000pt);
\draw (312.000000, 0.000000) node {\textcolor{white}{+}};
\end{scope}
\filldraw (312.000000, 15.000000) circle(1.500000pt);
\begin{scope}
\draw[fill=tq] (336.000000, 0.000000) circle(6.000000pt);
\clip (336.000000, 0.000000) circle(6.000000pt);
\draw (336.000000, 0.000000) node {\textcolor{white}{Z}};
\end{scope}
\draw (336.000000,60.000000) -- (336.000000,15.000000);
\begin{scope}
\draw[fill=tq] (336.000000, 60.000000) circle(6.000000pt);
\clip (336.000000, 60.000000) circle(6.000000pt);
\draw (336.000000, 60.000000) node {\textcolor{white}{+}};
\end{scope}
\filldraw (336.000000, 15.000000) circle(1.500000pt);
\begin{scope}
\draw[fill=tq] (360.000000, -0.000000) +(-45.000000:8.485281pt and 8.485281pt) -- +(45.000000:8.485281pt and 8.485281pt) -- +(135.000000:8.485281pt and 8.485281pt) -- +(225.000000:8.485281pt and 8.485281pt) -- cycle;
\clip (360.000000, -0.000000) +(-45.000000:8.485281pt and 8.485281pt) -- +(45.000000:8.485281pt and 8.485281pt) -- +(135.000000:8.485281pt and 8.485281pt) -- +(225.000000:8.485281pt and 8.485281pt) -- cycle;
\draw (360.000000, -0.000000) node {\textcolor{white}{H}};
\end{scope}
\end{tikzpicture}}
    \label{sfig:doble_opt_hy}
}
\caption{Optimized compilation of two-electron excitations to single qubit Rotation and \textsc{CNOT}: \subref{sfig:opt_jw} Full JW $\phi_{0\downarrow}\phi_{2\uparrow}\rightarrow\phi_{4\downarrow}\phi_{4\uparrow}$ \subref{sfig:doble_opt_hy} Hybrid JW-HCB with spatial-orbitals 0 and 2 encoeded in JW, and 1,3 encoded in HCB (marked with red wires) $\phi_{0\downarrow}\phi_{2\uparrow}\rightarrow\phi_{4\downarrow}\phi_{4\uparrow}$. $R_z$ gates marked in pink are carrying the gate parameter $\theta$.}
\label{fig:Compiled_Dobles_Opt}
\end{figure*}

\section{Applications}
\label{sec:application}
\subsection{Hybrid Encoding Error}
\label{subsec:HEE}
\begin{figure*}
\centering
\subfigure[]{
        \centering
    \includegraphics[width=0.4\textwidth]{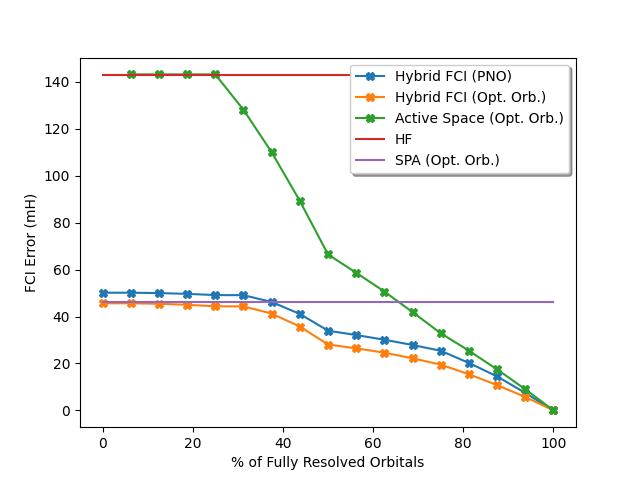}
    \label{sfig:pno_ch412} 
}
\subfigure[]{
        \centering
    \includegraphics[width=0.4\textwidth]{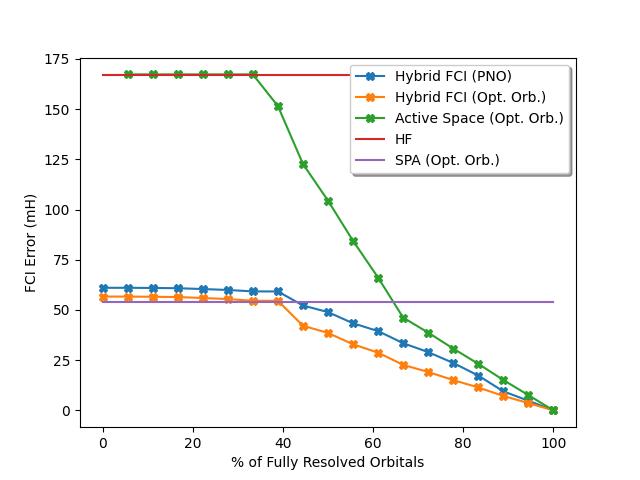}
    \label{sfig:pno_ethene12}
}
\caption{Hybrid-FCI error with respect to the exact diagonalization in all orbitals (FCI). \subref{sfig:pno_ch412} $CH_4$ with $n_{PNO}=16$. \subref{sfig:pno_ethene12} Ethene ($C_2H_4$) with $n_{orb}=18$. Both molecules use a basis of directly determined PNOs~\cite{kottmann2020reducing}. SPA corresponds to the separable pair approximation~\cite{kottmann2022optimized} operating on all spatial orbitals in HCB representation. Method/(Opt. Orb) refers to the usage of orbitals optimized (unitary rotations of the PNOs) to the SPA wavefunction. "Hybrid FCI (PNO)" and "Hybrid FCI (Opt. Orb.)" correspond to the Hybrid FCI at given \% of Fermionic Orbitals. "Active Space (Opt. Orb.)" corresponds to regular FCI  given \% of active space orbitals with respect to the complete basis FCI. }\label{fig:PNO_results}
\label{fig: PNO_results}
 \end{figure*}

Beyond reductions in circuit deepness, this approximation relies on a dimension compression through the usage of less qubits in the HCB registers, which will always have an associated error. In order to estimate this error, some methods have been proposed in the literature that could be adapted to this problem.~\cite{kim2024Variational} In the following, we will estimate it through the difference between a classical ``hybrid-FCI'', a FCI calculation on a Hamiltonian where all the integrals not allowed by the encoding are set to zero, and the ground state of the corresponding full-Fermionic Hamiltonian computed via FCI. To simultaneously accelerate Full Configuration Interaction (FCI) convergence and maintain precise spatial property representation, Pair Natural Orbitals (PNOs)~\cite{kottmann2020direct} were utilized. Figure~\ref{fig:PNO_results} shows the energy error with respect to the full Fermionic FCI, where three different series are represented. Two sets of calculations were performed using the hybrid FCI method. The first set, labeled ``Hybrid FCI (PNO)'' and ``Hybrid FCI (Opt. Orb.)'', employed either standard PNOs or PNOs re-optimized to a quantum circuit in the HCB encoding (here we used the separable pair approximation (SPA)~\cite{kottmann2022optimized} which has convenient properties for practical applications) level, respectively. In both cases, a fraction $x$ of the orbitals were treated as Fermionic within the hybrid FCI framework, while the remaining orbitals were considered Bosonic. Since PNOs are ordered by their occupation numbers, we have a relatively trustworthy heuristic to pick the right order of orbitals as they are sequentially resolved in the full-JW encoding.\\
The second set of calculations, termed ``Active Space (Opt. Orb.)'', involved full FCI computations over a restricted active space (using the same orbitals as SPA (Opt. Orb)) of the $x$\% lowest-energy orbitals obtained from the optimized orbital set (\textit{i.e.} all orbitals that are HCB encoded in the other calculations are here unoccupied and frozen). For both the SPA and HF energy calculations, the complete basis set was used, with orbital optimization performed for the SPA case.
\\\\ Results show that the Hybrid Energy Error (HEE) is always smaller than the Active Space restriction for the same \% of orbital restricted. Moreover, a relatively high active space is required to achieve the same error as a full Bosonic picture. Therefore, choosing a hybrid encoding with a shorter Fermionic sub-space, over the active-space restriction, is particularly suitable for scenarios demanding computational efficiency.
\subsection{Adaptive Encoding}
In order to illustrate how this hybrid encoding can be flexibly adapted to the problem of study, an ADAPT-based~\cite{grimsley2019adaptive} algorithm has been designed. For a given system, it aims to choose the most reduced Fermionic encoding without high energy loss for a given ansatz. This procedure allows to automatize the encoding selection. \\
\begin{figure}
    \centering
    \includegraphics[width=0.4\textwidth]{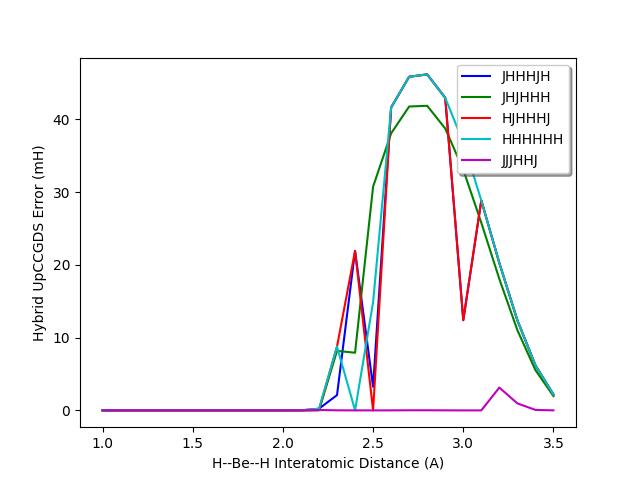}
    \caption{$BeH_2$ UpCCGSD HEE with respect to the Full Fermionic ansatz}
    \label{fig:adap_beh2}
\end{figure}
\begin{figure*}
    \centering
    \includegraphics[width=0.4\textwidth]{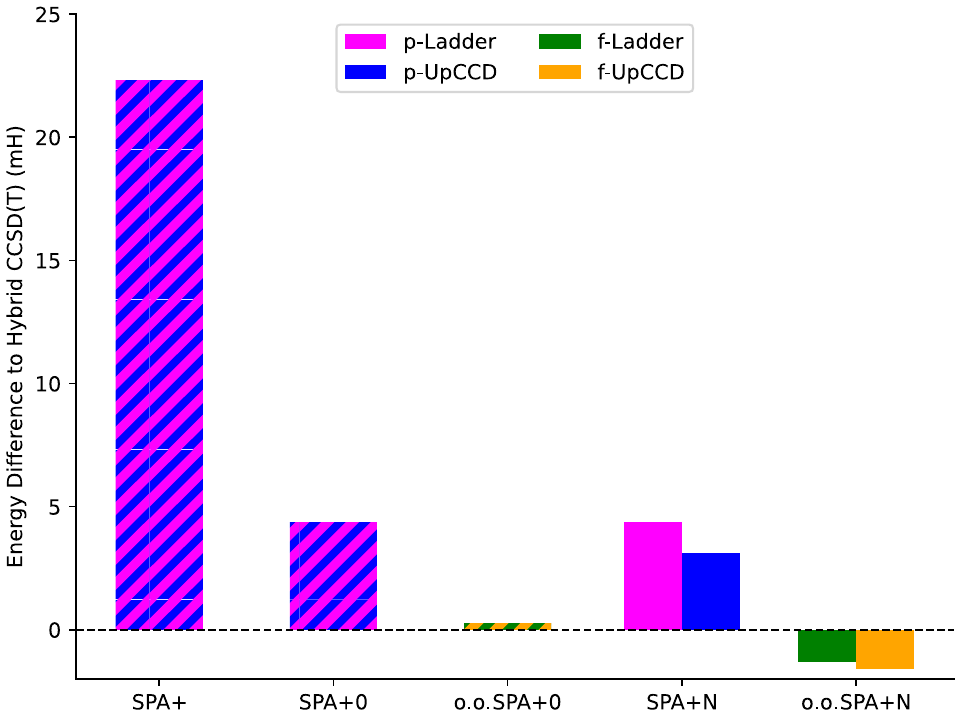}
    \raisebox{1cm}{\scalebox{0.65}{\input{Figures/SPA_circuit/opt_SPA_extended}}}

    \caption{Circuits Energy Errors with respect to the ``hybrid'' CCSD(T) energy.  Orbitals are optimized at SPA level. Results labeled with ``p-'' are obtained with orbital optimization restricted to the $\pi$ system while ``f-'' labels full optimization. ``SPA+'' correspond to the initial SPA circuit plus all elements on the $\pi$ system (identical circuit as  in~\cite{kottmann2023Molecular}).'SPA+0' corresponds to extending the base SPA+ by an UpCCD layer between the occupied and virtual Bosonic subspaces, and 'SPA+N' extend it by N sequences of correlators. These correlators are built in the ``Simple'' series by a simple ladder of excitations starting from the closest Fermionic orbital. On the other side, all the possible 2$e^-$ paired excitations are considered between the Fermionic and Bosonic subspaces for the ``UpCCD'' series.}
    \label{fig:but-circ}
\end{figure*}
For this purpose, the molecule will be initialized and optimized with some $U_{pre}$, in this case, the Hartree Fock state. The next step comprises a series of 1-step ADAPT cycles within a given pool. At each step, the molecular orbitals may be optimized to ensure better gradients and general results.~\cite{fitzpatrick2024SelfConsistent} At the end, the final circuit is analyzed selecting as Fermionic the orbitals affected by single or unpaired excitations and Bosonic otherwise. As an illustrative example, the $BeH_2$ symmetric di-dissociation curves with an UpCCGSD ADAPT pool have been simulated. \\\\Figure~\ref{fig:adap_beh2} shows the encoding error dissociation curve for the $BeH_2$ compared to the full Fermionic ansatz. Two main regimes can be identified for this system. Near the equilibrium, the molecule can be well described in a full-Bosonic way, and therefore all the encodings provide an energy close to the full Fermionic one. Conversely, an accurate description requires the inclusion of single excitations during dissociation (see~\cite{kottmann2023Compact,kottmann2023Molecular} for an analysis). \\\\The orbitals 1-6 in the employed basis (STO-3G) are 1: $\sigma_{BeH_2}$, 2: $\sigma_{BeH_2}^\dagger$, 3: $\pi_{BeH_2}$, 4:  Be $2p_y$, 5: $2p_{x}$ and 6: $\pi_{BeH2}$. Where the $\sigma_{BeH_2}$ MO is a linear combination of the \textsc{1s} H orbitals with the Be \textsc{2s}, and the $\pi_{BeH_2}$ is the linear combination of the Be \textsc{$p_z$} with each H \textsc{1s} on opposite phase. It is therefore reasonable that a comprehensive description of this dissociation necessitates the consideration of both the sigma and the two $\pi$ MOs. In contrast, the perpendicular $p_{x/y}$ orbitals do not significantly influence the outcome of this process. It could be argued that these results could have been predicted with some chemical intuition. We can assume here, that with a bit of chemical intuition, one can optimize the encoding selection to align with the specific problem of interest with relative ease.

\subsection{Circuits Designs}

In the previous section, the encoding was adapted to prominent methods like UpCCGSD and ADAPT. However, one can also do the opposite, design circuits with predefined subspaces in order to find a reasonable balance. As an illustrative example, we are going to employ the SPA+ circuit from~\cite{kottmann2023Molecular} in full Fermionic encoding within a minimal orbital basis and extend it into a larger set of orbitals represented in HCB encodings. The idea is to take the circuit as a base acting on a Fermionic subspace and transfer the electronic information to the Bosonic space while trying to maintain a low circuit depth. For this example, the Butadiene ($C_4H_6$) \textsc{sto-3g} has been chosen since it's already a system big enough to present some interest and it would be expensive to simulate any circuit in a full Fermionic schema. Due to the considerable system size, \textsc{FCI} becomes already inconvenient, therefore the reference energy is chosen to be \textsc{CCSD(T)}. This reference was tested to provide consistent results with (variational) CI methods such as \textsc{CISDT} and \textsc{CISDTQ} -- note that size consistency can be neglected here, as we are comparing absolute energies of a single instance.\\\\
We consider the 4 $p_z$ orbitals as Fermionic (forming the $\pi$ system) and all the other MOs Bosonic. Two \textsc{SPA+} circuit are contructed, both with the $\pi$ system identical to~\cite{kottmann2023Molecular}, where the $\pi$-system was chosen as active space, on the Fermionic subspace, optimizing the $\pi$ orbitals for the SPA. On the 'p-' series, the bosonic orbitals al left untouched, optimizing only the fermionic subspace, while on the 'f-' series, the SPA and orbital optimization is extended for all the Bosonic subspaces. Then,  a further initialization of the Bosonic subspaces is done with an UpCCD ansatz. At this point we have a product state between the SPA+ on the Fermionic and the UpCCD on the Bosonic registers. To correlate the two registers, different sets of correlators (two-electron paired excitation gates) are added between the Fermionic and Bosonic subspaces. \\ \\ 
Figure~\ref{fig:but-circ} shows the energy errors relative to the hybrid classical ground state energy with canonical orbitals, for the previously described SPA+ and SPA+N circuits. The hybrid energy is calculated in a manner analogous to the approach delineated in Subsection~\ref{subsec:HEE},  but employing the CCSD(T) method. Two classes of correlators were constructed to explore various approaches. The "simple" class comprises correlators progressing from a single fermionic orbital to Bosonic orbitals assembled in ladder form. In contrast, UpCCD incorporates all double excitations between both subspaces. The reported circuits are progressively built by first introducing UpCCD only between the orbitals belonging to the Bosonic occupied orbitals to the Bosonic virtual ones (SPA+0), and subsequently adding "Correlator N" (SPA+N).
\\\\These results show that a large part of the energy error comes from the hybrid encoding error, which can be addressed to the small Fermionic space. Then, most of the remaining error can be recovered by the Bosonic occupied-virtual UpCCD, which can be built in parallel with the first \textsc{SPA} graph and can be compiled to qubit excitations, reducing considerably the total circuit depth compared to a potential full Fermionic equivalent. Later, the additional correlators recover part of the missing unidentified energy, but around 3 mH are still missing with this approach. Further results suggest that increasing the initial SPA circuit in order to allow mixing the \textit{occupied} and \textit{virtual} orbitals, followed by consequent orbital relaxation, presented improved results at the range of $\pm$1 mH energy error already for the SPA+0 circuits. Latter mixing between Fermionic and Bosonic subspaces show a small energy reduction, but not too much else may be recovered for this system.
\\Within this work, we limit the scope to a technical illustration of how one could take advantage of this Hybrid Encoding to design circuits which may go further than what is feasible on nowadays NISQ computers.

\section{Conclusion \& Outlook}
In this work, we developed a hybrid Fermion-to-qubit encoding, which allows us to split the molecular Fock space into Fermionic and quasi-Bosonic parts. We have illustrated the main capabilities and how resources like qubit count, circuit depth and the amount of measurements can be reduced. However, it is important to keep in mind that this approximation comes at a cost, an error associated with the lack of flexibility of the wave function. This error will furthermore depend on the used orbital basis -- here we have used pair-natural orbitals~\cite{kottmann2020reducing} as well as optimized-orbitals adapted to an SPA circuit as an initial demonstration. Preliminary results suggest that further orbital-optimization can be advantageous in finding the optimal representation of the chosen encoding. Further research along those lines is currently being conducted. 
\section*{Conflicts of interest}
There are no conflicts to declare.

\section*{Scientific Software and Code Availability}
Development and data generation within this work has been conducted trough the open-source package \textsc{tequila}~\cite{tequila} using \textsc{qulacs}~\cite{qulacs} as simulation backend, the JW transformation from \textsc{open-fermion}~\cite{OpenFermion}, molecular integrals as well as classical methodology from \textsc{pyscf}~\cite{pyscf1,pyscf2, pyscf3} and the automatically differentiable framework described in Ref.~\cite{kottmann2021feasible}. PNOs where computed with \textsc{madness}~\cite{harrison2016madness} using methods described in Refs.~\cite{kottmann2020direct, kottmann2020reducing}. Circuits are created via \textsc{qpic}~\cite{qpic}.\\
An implementation of the developed encodings will be made available as an extensions to \textsc{tequila}. We provide initial examples in a \href{https://github.com/JdelArco98/Quantum_Chemistry_Hybrid_Base}{github repository}.~\cite{repo}

\section*{Acknowledgement}
This work has been funded by the Hightech Agenda Bayern and the Munich Quantum Valley through the Lighthouse project KID-QC$^2$. We thank Davide Bincoletto for various discussions, as well as Johannes T\"olle for insights regarding future projects. The authors gratefully acknowledge the resources on the LiCCA HPC cluster of the University of Augsburg, co-funded by the Deutsche Forschungsgemeinschaft (DFG, German Research Foundation) – Project-ID 499211671.

\label{app:circuits}

\bibliography{main.bib}
\end{document}